\documentclass[aps, twocolumn, prl, reprint, notitlepage, floats, floatfix, amsmath, amssymb, amsfonts, superscriptaddress, showpacs, showkeys, nofootinbib]{revtex4-2}
\usepackage{amsmath,amsfonts,amssymb,amsthm,bm}
\usepackage{graphicx}
\usepackage[usenames,dvipsnames]{xcolor}
\usepackage[utf8]{inputenc} 
\usepackage[
	bookmarksnumbered, bookmarksopen, bookmarksopenlevel=2,
	breaklinks=true,
	colorlinks=true, filecolor=mylinkcolor, citecolor=mylinkcolor,
	linkcolor=mylinkcolor, urlcolor=mylinkcolor, menucolor=mylinkcolor,
]{hyperref}
\usepackage{booktabs}
\usepackage{multirow}
\xdefinecolor{mylinkcolor}{rgb}{0,0,0.5}

\newcommand{\LdIT}{\affiliation{Laboratoire des 2 Infinis - Toulouse (L2IT-IN2P3), Université de Toulouse, CNRS, UPS, F-31062 Toulouse Cedex 9, France}}
\newcommand{\UZH}{\affiliation{Physik-Institut, Universät Zürich, Winterthurerstrasse 190,
8057 Zürich, Switzerland}}

\newcommand{\SISSA}{\affiliation{SISSA, Via Bonomea 265, 34136 Trieste, Italy and INFN Sezione di Trieste}}
\newcommand{\Pisa}{\affiliation{Dipartimento di Fisica ``Enrico Fermi'', Universit\`a di Pisa, Largo Pontecorvo 3, I-56127 Pisa, Italy}}
\newcommand{\INFNp}{\affiliation{INFN, Sezione di Pisa, Largo Pontecorvo 3, I-56127 Pisa, Italy}}
\newcommand{\AEI}{\affiliation{Max-Planck-Institut f\"ur Gravitationsphysik, Albert-Einstein-Institut, Am M\"uhlenberg 1, 14476 Potsdam-Golm, Germany}}
\newcommand{\MB}{\affiliation{Dipartimento di Fisica ``G. Occhialini'', Universit\`a di Milano - Bicocca, Piazza della Scienza 3, 20126 Milano, Italy}}
\newcommand{\INFNm}{\affiliation{INFN, Sezione di Milano-Bicocca, Piazza della Scienza 3, 20126 Milano, Italy}}
\newcommand{\CERN}{\affiliation{Theoretical Physics Department, CERN, 1211 Geneva 23, Switzerland}}
\newcommand{\Gen}{\affiliation{Département de Physique Théorique and Center for Astroparticle Physics, Université de Genève, Quai E. Ansermet 24, 1211 Genève 4, Switzerland}}
\newcommand{\ifpu}{\affiliation{IFPU - Institute for Fundamental Physics of the Universe, Via Beirut 2, 34014 Trieste, Italy}}

\begin{document}

\title{Reducing cosmological degeneracies by combining multiple classes of\\LISA gravitational-wave standard sirens}
\author{Danny Laghi}
\email{danny.laghi@physik.uzh.ch}\UZH\LdIT
\author{Nicola Tamanini}\LdIT
\author{Alberto Sesana}\MB\INFNm
\author{Jonathan Gair}\AEI
\author{Enrico Barausse}\SISSA\ifpu
\author{Chiara Caprini}\CERN\Gen
\author{Walter Del Pozzo}\Pisa\INFNp
\author{Alberto Mangiagli}\AEI
\author{Sylvain Marsat}\LdIT

\begin{abstract}
We present the first joint gravitational-wave cosmological inference with LISA extreme mass-ratio inspirals at $z\lesssim1$ (galaxy redshifts) and massive black hole binaries at $z\gtrsim1$ (electromagnetic counterparts). Combining these standard sirens reduces cosmological degeneracies and yields competitive constraints on the Hubble constant $H_0$ and the dark-energy equation-of-state parameter $w_0$. This highlights LISA's potential for late-time cosmology across a broad redshift range with systematics distinct from electromagnetic distance indicators.
\end{abstract}

\maketitle

\noindent \textit{Introduction---}
Over the past decade, a series of increasingly precise cosmological measurements has begun to place the flat $\Lambda$CDM concordance model under growing tension---despite its remarkable empirical success~\cite{Turner:2018bcg,perivolaropoulos2022challenges}.  
A prominent example is the persistent discrepancy between early- and late-Universe determinations of the present-day expansion rate, the Hubble constant $H_0$~\cite{Planck:2018vyg,Riess:2021jrx,Kamionkowski:2022pkx}. In parallel, recent large-scale structure results have been interpreted as possible evidence for departures from a cosmological-constant description of dark energy (DE), motivating tests of dynamical DE scenarios~\cite{DESI:2024mwx,DESI:2025zgx}. 

Compact binaries emitting gravitational waves (GWs)  constitute standard cosmological rulers, commonly called \emph{standard sirens}~\cite{Holz:2005df}, offering a promising independent way to chart the cosmic expansion history across redshifts~\cite{Schutz:1986gp}. 
In fact, the luminosity distance can be directly derived from observations, while the binary's redshift can be estimated through several techniques, including an electromagnetic (EM) counterpart (\emph{bright sirens})~\cite{Tamanini:2016zlh,LIGOScientific:2017adf,Mangiagli:2022niy}, 
or sharp features in the binary population mass distribution (\emph{spectral sirens})~\cite{Chernoff:1993th,Taylor:2011fs,Farr:2019twy,Mastrogiovanni:2021wsd,Ezquiaga:2022zkx}, which can be complemented by the use of galaxy catalogs to cross-match with the GW sky localization (\emph{dark sirens})~\cite{MacLeod:2007jd,Laghi:2021pqk,Zhu:2021aat,LIGOScientific:2021aug,Liu:2025sgz,LIGOScientific:2025jau}. Other methods include the spatial cross-correlation of GWs and galaxies or GW weak lensing~\cite{Camera:2013xfa,Oguri:2016dgk,Mukherjee:2019wcg,Mukherjee:2020hyn,Mukherjee:2022afz,Fonseca:2023uay,Zazzera:2024agl,Ferri:2024amc,Pedrotti:2025tfg, Mpetha:2022xqo} 
and tidal effects in neutron stars~\cite{Messenger:2011gi,DelPozzo:2015bna,Chatterjee:2021xrm}.

The Laser Interferometer Space Antenna (LISA)~\cite{LISA:2017pwj,LISA:2024hlh}, a space mission adopted by the European Space Agency in 2024 with launch planned for the mid-2030s, will for the first time observe GWs in the mHz band. 
The mission will deliver at least 4.5 years of data, with a possible extension to 10 years~\cite{LISA:2024hlh}.

LISA will observe extreme-mass-ratio inspirals (EMRIs)~\cite{Amaro-Seoane:2012lgq,Babak:2017tow}, in which a stellar-mass compact object (black hole, neutron star, or white dwarf) gradually spirals into a massive black hole (MBH)of mass $10^5$--$10^7 M_{\odot}$. 
Besides EMRIs, LISA will detect the inspiral-merger-ringdown signal of massive black hole binaries (MBHBs)~\cite{Klein:2015hvg,Bonetti:2018tpf,Barausse:2020mdt,Barausse:2023yrx}, with total mass $10^4$--$10^7 M_{\odot}$, and the inspiral-only phase of binaries with mass $10^3$--$10^4 M_{\odot}$~\cite{LISA:2024hlh}. These systems may produce EM counterparts in gas-rich environments~\cite{Tamanini:2016zlh,Bogdanovic:2021aav,Mangiagli:2022niy}.
One of LISA's primary science objectives is to probe the cosmic expansion~\cite{LISA:2024hlh,LISACosmologyWorkingGroup:2022jok}.
EMRIs can serve as dark sirens to at low to intermediate redshift~\cite{MacLeod:2007jd,Laghi:2021pqk,LISACosmologyWorkingGroup:2022jok,liu2024probing,Lyu:2024gnk},
while MBHBs with identified cunterparts are bright sirens enabling high-redshift constraints~\cite{Tamanini:2016zlh,Caprini:2016qxs,Belgacem:2019tbw,Speri:2020hwc,Mangiagli:2023ize}.

In this article, we present a new analysis of EMRI and MBHB standard sirens and quantitatively combine, for the first time, LISA dark and bright sirens within a joint Bayesian framework, showing that their complementarity partially breaks the degeneracy among late-time $\Lambda$CDM parameters, leading to improved parameter constraints. This also offers improved constraints on the DE parameter, $w_0$, in a dynamical DE scenario.
Our results suggest that LISA will become a unique cosmological probe able to map the expansion of the Universe from the local universe out to high redshift ($z\gtrsim 5$)~\cite{Tamanini:2016uin}.\\

\textit{Methods---} 
We model cosmological inference in a hierarchical Bayesian approach~\cite{Mandel:2018mve}, treating dark and bright sirens separately.
We consider a set of GW events $\{D\} = \{D_i\}^N_{{i=1}}$, where $N$ is the number of detected EMRIs or MBHBs passing a selection threshold in GW signal-to-noise ratio (SNR), with the SNR a deterministic function of the data. 
For each event, we summarize $D_i$ by LISA measured sky region $\Delta \hat{\Omega}$ and luminosity distance $\hat{d}_L$ (with instrumental and weak lensing contributions), using the single-event likelihood marginalized over all other source parameters. 
We are interested in the posterior distribution for the cosmological parameter vector $\lambda$, which, assuming events are individually resolved, is given by~\cite{Mandel:2018mve,Laghi:2021pqk,Gair:2022zsa}
\begin{equation}\label{eq:posterior}
\begin{split}
    p(\lambda | \{D\}) \propto \; & \frac{p(\lambda)}{\alpha(\lambda)^N} \times \prod_{i=1}^N  \int \! dz \int d {\Omega} \; \\
    & \times \mathcal{L}_{\rm GW}(D_i | d_L(z, \lambda), {\Omega}) \, p(z, {\Omega}| \lambda) \text{.}
\end{split}
\end{equation}
A detailed derivation and explicit expressions for the different terms in Eq.~\eqref{eq:posterior} are provided in the End Matter. The parameters $z$ and ${\Omega}$ are the GW source redshift and angular sky location, respectively, while $\mathcal{L}_{\rm GW}(D_i | d_L(z, \lambda), {\Omega})$ is the GW likelihood, representing the probability of getting the data $D_i$ given a GW angular sky location ${\Omega}$, and a luminosity distance $d_L(z, \lambda)$. The latter, assuming spatial flatness at cosmological scales, can be generally expressed as
\begin{equation}
d_L(z, \lambda) = c(1+z) \int_0^z \frac{d \tilde{z}}{H(\tilde{z})}\,,
\label{eq:dL_z}
\end{equation}
where
\begin{equation}
\begin{split}
H(z) &= H_0 \sqrt{\Omega_{m}(1+z)^3+(1-\Omega_{m}) f(z)}\,,\\
f(z) &= (1+z)^{3(1+w_0+w_a)}e^{-3 w_a \frac{z}{1+z}}\,.
\end{split}
\label{eq:Hubble_parameter}
\end{equation}
Here $H_0$ is the value of the Hubble constant today, $\Omega_{m}$ is the present fraction of matter energy density, $c$ is the speed of light, and $w_0$ and $w_a$ are DE equation-of-state parameters defined by $w(z) = w_0 + w_a z/(1+z)$~\cite{Chevallier:2000qy,Linder:2002et}, with the $\Lambda$CDM model recovered when $(w_0,w_a)=(-1,0)$.
The term $p(z, {\Omega} | \lambda)$ is the GW redshift and sky location prior, which (potentially) depends on the cosmological parameters $\lambda$. This prior represents the redshift information coming from potential host galaxies in a galaxy catalog or, alternatively, from an identified EM counterpart, 
in a dark and bright siren analysis, respectively. The term $p(\lambda)$ is the prior distribution for the cosmological parameters, while $\alpha(\lambda)$ accounts for GW selection effects arising from the finite detector sensitivity.

A fully hierarchical standard siren analysis requires joint inference of cosmological and population parameters~\cite{Mastrogiovanni:2023emh,Gray:2023wgj,Borghi:2023opd,LIGOScientific:2025jau}. 
In general, both dark and bright siren analyses rely on a model for the distribution of source parameters correlated with $\lambda$, implying that $\alpha(\lambda)$ depends on both cosmology and population. Since fully self-consistent modeling of EMRI and MBHB populations and selection effects for cosmology remains computationally challenging 
(see~\cite{Chapman-Bird:2022tvu,Singh:2026sad} for related approaches in the EMRI context), we perform inference only in the cosmological parameter space and include selection effects via $\alpha(\lambda)$, evaluated under the assumed source population (see End Matter for details).

The redshift and sky location prior, $p(z, \Omega | \lambda)$, is constructed differently for dark and bright sirens; we summarize the procedure here and refer to the End Matter for details.

For dark sirens, we cross-match a galaxy catalog with the GW 3D localization volume in redshift space~\cite{Laghi:2021pqk,Muttoni:2021veo,muttoni2023,liu2024probing}. 
For each event, we retain galaxies within the measured 3$\sigma$ GW sky region, $\Delta \hat{\Omega}$, and within the largest redshift shell mapped from the 3$\sigma$ luminosity distance interval via Eq.~\eqref{eq:dL_z} and permitted by $p(\lambda)$, accounting for additional redshift uncertainty due to galaxy peculiar velocities. 
We remain agnostic about the astrophysical properties of the host galaxies and assign an equal probability to each galaxy of hosting the GW event. The galaxy redshift distribution is modeled with a Gaussian mixture that accounts for the location of the galaxies in the sky relative to the EMRIs, yielding an event-specific interpolant for $p(z, \Omega | \lambda)$ used in Eq.~\eqref{eq:posterior}.

For bright sirens, we 
assume a uniquely identified galaxy host and take $p(z, {\Omega} | \lambda)$ to be the host redshift posterior (Gaussian) together with the measured sky location. We adopt spectroscopic redshift uncertainty $\sigma_z = 10^{-3}$, or photometric uncertainties $\sigma_z=0.2$ and 0.5 for Lyman-$\alpha$ and Balmer-break determinations, respectively~\cite{Mangiagli:2022niy}.

The posterior distributions, $p(\lambda|\{D\})$, from the dark and bright siren analyses are finally combined to obtain a joint posterior distribution, from which the marginal distributions of the cosmological parameters can be derived.\\

\textit{Case study---}
We consider two cosmological scenarios: (i) Flat $\Lambda$CDM with $\lambda = \{ h,\Omega_{m} \}$, where $h \equiv H_0/100 \,\text{km}^{-1}\,\text{s}\,\text{Mpc}$, uniform priors $h\in[0.6,0.76]$ and $\Omega_{m} \in [0.04, 0.5]$, and fiducial values $(h,\Omega_m) = (0.673, 0.315)$ matching the cosmology of the galaxy catalog used for the dark-siren analysis; (ii) Dynamical DE with $\lambda = \{w_0, w_a\}$, assuming $(h,\Omega_m)$ fixed by external probes, uniform priors $w_0 \in [-3, -0.3]$, $w_a \in [-1, 1]$, and fiducial $(w_0, w_a) = (-1, 0)$ corresponding to the cosmological constant $\Lambda$. 
For $\Lambda$CDM we also perform 1D runs fixing either $h$ or $\Omega_m$ to its fiducial value.

We apply our method to state-of-the-art simulated EMRI and MBHB catalogs.
The GW EMRI data come from~\cite{Babak:2017tow}, adopting their \emph{fiducial} population model \emph{M1} and Fisher information matrix (FIM) parameter uncertainties computed with the analytic-kludge ``Schwarzschild'' (AKS) waveform~\cite{Barack:2003fp}. Within that framework, AKS is more conservative than more optimistic variants considered in ~\cite{Babak:2017tow} in terms of SNR and detection rates; we do not explore alternative population scenarios (see~\cite{Babak:2017tow}).
The MBH masses are drawn from a mass function based on the \emph{Model PopIII}~\cite{Madau:2001sc,Volonteri:2002vz,Klein:2015hvg}, a self-consistent model for MBH formation and cosmic evolution~\cite{Barausse:2012fy,Sesana:2014bea,Antonini:2015cqa,Antonini:2015sza}, where MBHs grow from the remnants of metal-poor population-III stars (light seeds).
We retain EMRIs with SNR$>$50. This high-SNR threshold guarantees the inclusion of highly-informative, well-localized events, while keeping the analysis's computational cost feasible. In this sample, 68\% of the events have 90\% credible interval (CI) sky localization $\Delta \hat{\Omega}_{90}/\text{deg}^2 < 2$ (min/median/max: $6\cdot10^{-3}$ deg$^2$, 1.1 deg$^2$, 48 deg$^2$).

For dark sirens we cross-match EMRIs to a simulated galaxy catalog following the steps detailed in the End Matter. We use a customized light cone~\cite{Henriques:2014sga,IzquierdoVillalba2019,muttoni2023} covering one octant to $z<1$. We assume completeness above $M_{*} > {10^{10}}M_{\odot}$ over this redshift range, as a representative Stage V scenario (e.g., MUST, Spec-S5, WST)~\cite{Zhao:2024alp,Spec-S5:2025uom,WST:2024rai}; relaxing this assumption can be incorporated through a completeness/selection function~\cite{Borghi:2025pav,Moore:2025mwj}.

For the MBHB bright siren catalogs (with counterparts), 
we rely on the simulated catalogs presented and described in~\cite{Mangiagli:2022niy}. To ensure population consistency with the EMRI model, we adopt the MBHB astrophysical model labeled \emph{PopIII} in~\cite{Barausse:2012fy,Klein:2015hvg}, since this is based on the same MBH population model used in the EMRI model M1. This model yields the most conservative number of EM counterpart detections among the MBHB population models considered in~\cite{Mangiagli:2022niy}. GW parameters and FIM  uncertainties are obtained with the waveform model \textsc{IMRPhenomHM}~\cite{London:2017bcn}   

We assume that the optical and X-ray emissions are produced after the merger, while the radio emission is produced near the merger. Independently from the type of emission, the observation is performed post-merger because the sky localization estimates have been computed with the full inspiral-merger-ringdown waveform.
We explore three EM detection scenarios for the EM identification and redshift measurement, adopting the \emph{maximizing} scenario of~\cite{Mangiagli:2022niy}: (i) the Vera C. Rubin Observatory~\cite{LSST:2008ijt} alone (identification + redshift); (ii) SKA~\cite{Dewdney:2009tmd} (identification) + ELT~\cite{ELT} (redshift); (iii) Athena~\cite{Nandra:2013jka,Piro:2022zos} (identification) + ELT (redshift).
Following~\cite{Mangiagli:2022niy}, our MBHB data set consists of GW events with SNR$>$10 and Bayesian sky localization $\Delta\hat{\Omega}_{90} < 10 \text{ deg}^2$ or $\Delta\hat{\Omega}_{90} < 0.4 \text{ deg}^2$ to guarantee detection with the Vera Rubin Observatory and SKA, or with Athena, respectively.

We study two mission durations consistent with the baseline schedule~\cite{LISA:2024hlh}: 4 years (\emph{fiducial}), corresponding to an overall nominal mission duration of 4.25 years with an 82\% duty cycle for data collection (yielding 3.7 years, approximated to 4 years), and 10 years.

To increase the statistical robustness of our results, for each duration we generate 20 independent realizations: for EMRIs by repeating the EMRI-light cone cross-match detailed in the End Matter, and for MBHBs by resampling the counterpart catalogs as in~\cite{Mangiagli:2022niy}.
Table~\ref{tab:N_EMRI_MBHB} reports $\langle N \rangle$ for EMRIs and MBHBs averaged over realizations.\\

\begin{table}[t]
\centering
\caption{Summary table reporting the average number of sources, $\langle N\rangle$, and the ensemble-averaged fractional uncertainties $\langle \delta_x \rangle$, where $\delta_x \equiv (x_{84} - x_{16})/(2x_{50})$ and $x=\{h, \Omega_m, w_0 \}$, computed over 20 different data realizations for the $\Lambda$CDM and DE scenarios, for 4 and 10 years of the LISA mission. The columns under ``$\Lambda$CDM (1D)'' are obtained by repeating the analysis and setting one of the two parameters to its fiducial value. We do not report results for $w_a$ as this parameter is unconstrained with all sources.}
\label{tab:N_EMRI_MBHB}
\begin{tabular}{l c @{\hspace{16pt}} cc cc c}
\toprule
& $\langle N\rangle$
& \multicolumn{2}{c}{$\Lambda$CDM (2D)} 
& \multicolumn{2}{c}{$\Lambda$CDM (1D)} 
& DE (2D)\\
\cmidrule(lr){3-4}\cmidrule(lr){5-6}
Source
&
& $h [\%]$ & $\Omega_m [\%]$
& $h [\%]$ & $\Omega_m [\%]$
& $w_0 [\%]$ \\
\midrule

\multicolumn{7}{l}{\textbf{4 yr}}\\
\midrule
EMRI
& 60  & 2.0 & 25 & 0.6 & 7.9 & 9.3 \\
MBHB
& 6   & 4.3 & 19 & 1.3 & 6.5 & 14 \\
EMRI+MBHB
& 66  & 1.2 & 9.4 & 0.5 & 3.9 & 7.2 \\

\midrule
\multicolumn{7}{l}{\textbf{10 yr}}\\
\midrule
EMRI
& 156 & 1.0 & 8.6 & 0.2 & 1.5 & 5.9 \\
MBHB
& 16  & 1.5 & 9.1 & 0.5 & 3.2 & 7.6 \\
EMRI+MBHB
& 172 & 0.6 & 4.7 & 0.3 & 1.2 & 4.9 \\

\bottomrule
\end{tabular}
\end{table}

\textit{Results---}
We compute the posterior in Eq.~\eqref{eq:posterior} for 20 realizations of the dark and bright siren catalogs separately, and for both $\Lambda$CDM and DE scenarios.
Joint EMRI+MBHB constraints are obtained by randomly pairing dark and bright siren realizations and multiplying the corresponding likelihoods, yielding an ensemble of 20 joint posteriors for $\lambda$. 

For each realization we marginalize $p(\lambda|\{D\})$ to 1D posteriors and for each parameter $x$ compute the half-width of the central 68\% CI,  $\sigma_{x} \equiv (x_{84} - x_{16})/2$, where $x_{q}$ denotes the $q$-th posterior quantile, and the \emph{fractional uncertainty} $\delta_x \equiv \sigma_{x}/x_{50}$, with $x_{50}$ the posterior median. To capture realization-to-realization fluctuations, we report \emph{ensemble averages} $\langle\delta_x\rangle$ across the 20 realizations; this provides a compact summary of typical performance, though the scatter can be larger for MBHBs due to small-number fluctuations~\cite{Mangiagli:2022niy}.
Table~\ref{tab:N_EMRI_MBHB} summarized our results. 

In the fiducial 4-year LISA scenario, EMRIs constrain the Hubble constant more tightly than MBHBs (2.0\% vs 4.3\%), while MBHBs constrain $\Omega_m$ more effectively than EMRIs ($19\%$ vs $25\%$), albeit both remain relatively weak. 
As illustrated in Fig.~\ref{fig:LambdaCDM_regression}, this reflects: (i) the different redshift leverage of the two populations, as dark sirens dominantly probe $z\lesssim 1$ and are most sensitive to $H_0$, whereas bright sirens extend to $z\gtrsim 1$ and provide greater leverage on $\Omega_m$; and (ii) the smaller number of MBHBs relative to EMRIs in 4 years.

\begin{figure}[t]
    \centering
{\includegraphics[width=0.49\textwidth]{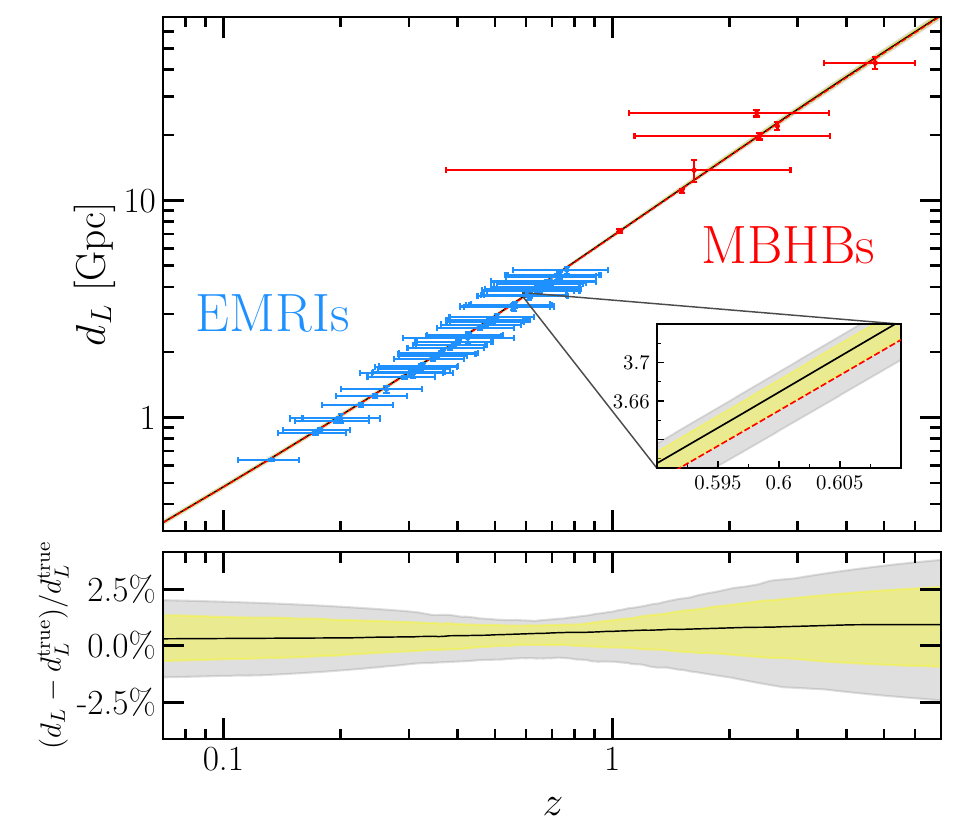}}
    \caption{Reconstructed $d_L$-$z$ regression line for the representative 4-year LISA detection scenario shown in Fig.~\ref{fig:LambdaCDM_2d} (see main text for more details): we show the median (solid black) and $68\%$ and $90\%$ credible regions in yellow and light gray, respectively. The red dashed line corresponds to the fiducial cosmology. Each data point shows the $1\sigma$ uncertainty of $d_L$ (including LISA instrumental and weak lensing uncertainties), while the redshift uncertainties correspond to the redshift shell over which each single-event likelihood is marginalized. For the MBHBs data points, larger (smaller) redshift error bars correspond to photometric (spectroscopic) follow-ups.
    The inset shows the most constrained redshift region, while the bottom panel shows the residuals of the inferred regression line and its credible regions.}
    \label{fig:LambdaCDM_regression}
\end{figure}
\begin{figure}[t]
    \centering
    \includegraphics[width=0.49\textwidth]{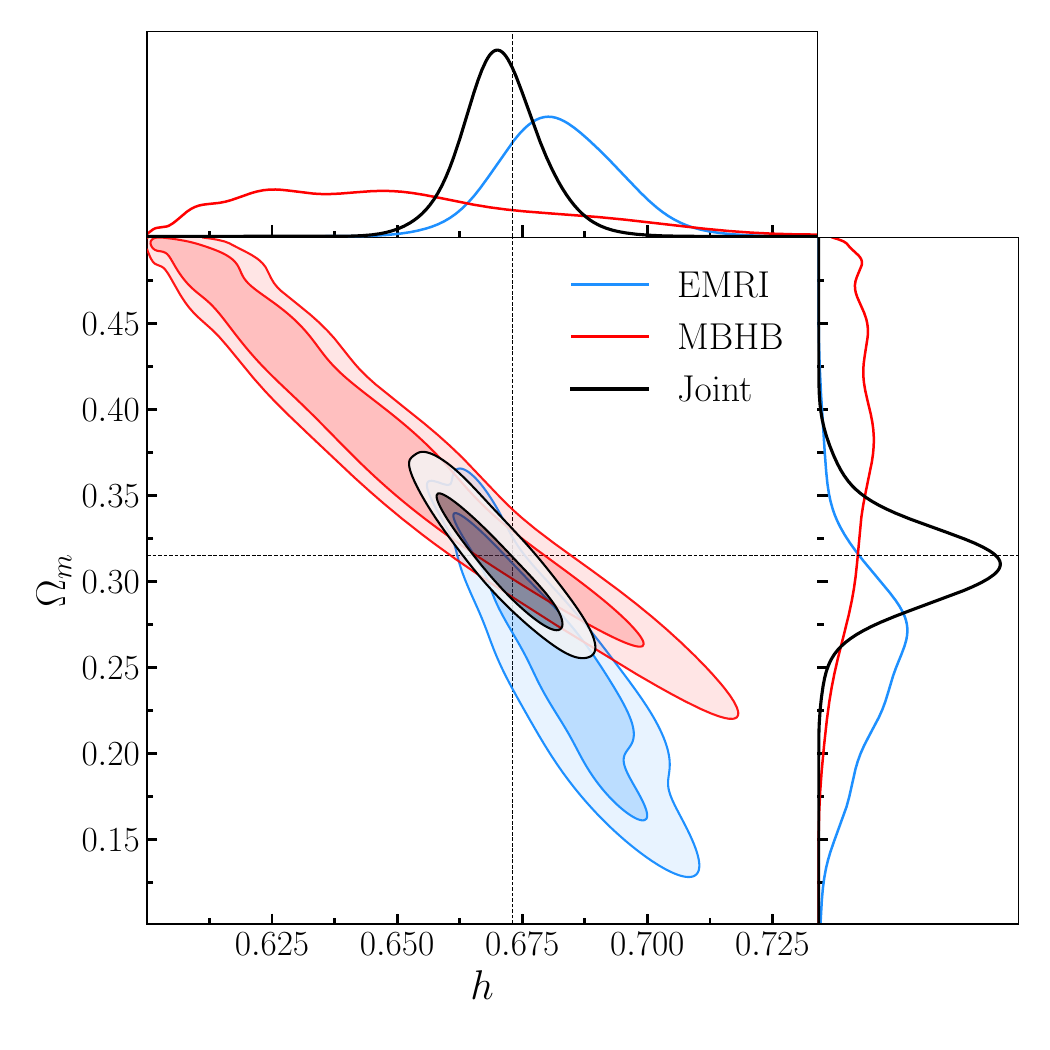}%
    \caption{Constraints on the two parameters $h$ and $\Omega_m$ in the $\Lambda$CDM model from EMRI dark sirens (blue), MBHB bright sirens (red), and their joint combination (black), for the representative 4-year realization shown in Fig.~\ref{fig:LambdaCDM_regression}. The contours show the 68\% and 90\% confidence levels, while the dashed lines show the fiducial cosmology.  The top and side panels show 1D marginalized constraints.}
    \label{fig:LambdaCDM_2d}
\end{figure}

Fixing one $\Lambda$CDM parameter to its fiducial value (e.g., by extremely precise CMB+BAO or other EM constraints) improves the other by a factor of $\sim$3 for both populations (Table~\ref{tab:N_EMRI_MBHB}). 
The $h$-only inference closely mirrors the strategy commonly employed in current dark-siren analyses based on galaxy catalogs~\cite{Mastrogiovanni:2023emh,Gray:2023wgj,LIGOScientific:2025jau}, where the limited redshift leverage of low-$z$ events motivates fixing $\Omega_m$ to obtain constraints on $H_0$. 

Turning to dynamical DE, sampling ($w_0$, $w_a$), both populations constrain $w_0$ (9.3\% for EMRIs; 14\% for MBHBs) but remain largely insensitive to $w_a$, as expected given the statistical and redshift leverage required to detect time evolution.

The full cosmological potential of LISA standard sirens emerges when dark- and bright-siren samples are combined. Owing to their distinct redshift distributions, the two populations exhibit complementary degeneracy directions in the $\Lambda$CDM cosmological parameter space: as one moves from predominantly low-$z$ to higher-$z$ sources, the principal degeneracy direction rotates (counter-clockwise in the $h$-$\Omega_m$ plane), analogous to the complementarity observed among EM probes that span different redshift ranges (see, e.g.,~\cite{DESI:2025zgx}). This complementarity is illustrated in Fig.~\ref{fig:LambdaCDM_2d}, where we show a representative realization selected by the median precision of the marginalized joint posterior on $h$.

The combined dark+bright analysis yields $\langle \delta_h \rangle \simeq 1.2\%$ and $\langle \delta_{\Omega_{m}} \rangle \simeq 9.4\%$. The constraint on $h$ is comparable in precision to recent Type Ia supernova determinations~\cite{Riess:2021jrx}, while the $\Omega_{m}$ constraints remain less competitive than the tightest CMB+BAO combinations~\cite{Planck:2018vyg}, but represents a substantial improvement over either LISA populations alone. 
Relative to the EMRI-only analysis, the improvement is $\sim$1.7 in $h$ and $\sim$2.6 in $\Omega_{m}$, and relative to MBHBs alone $\sim$3.6 and $\sim$2.0, respectively.
With one $\Lambda$CDM parameter fixed externally, constraints tighten to $\langle \delta_h \rangle \simeq 0.5\%$ (or $\langle \delta_{\Omega_{m}} \rangle \simeq 3.9\%$). 

Fig.~\ref{fig:DE_w0_1D} shows the joint posterior on $w_0$ for a representative realization, selected with the same criterion used for Figs.~\ref{fig:LambdaCDM_regression} and~\ref{fig:LambdaCDM_2d}, but applied to $w_0$. In the DE case, combining populations yields $\langle \delta_{w_0} \rangle \simeq 7.2\%$, an improvement of $\sim$1.3-1.9 over the individual populations. While weaker than current  tight multi-probe constraints~\cite{Planck:2018vyg,Brout:2022vxf}, that combine CMB information with late-time distance measurements ($\sim$2-3\%), it is substantially stronger than supernovae-only constraints ($\sim$10-16\%)~\cite{Brout:2022vxf}, providing an independent and complementary measurement with different systematics.

\begin{figure}
    \includegraphics[width = 0.49\textwidth]{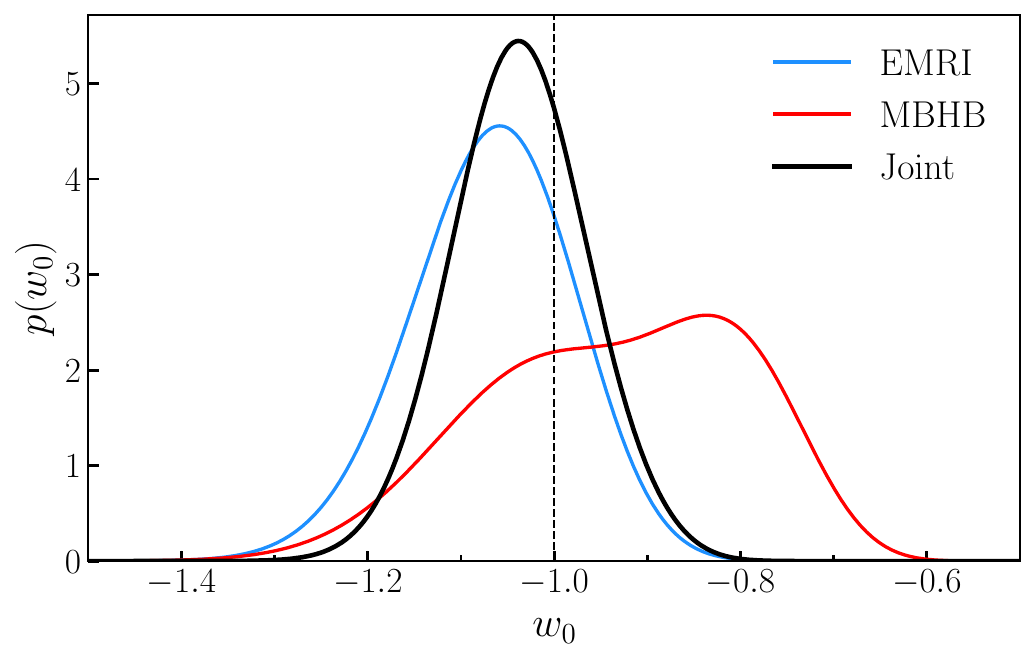}
    \caption{Constraints on the $w_0$ parameter in the DE scenario from EMRI dark sirens (blue), MBHB bright sirens (red), together with their joint combination (black), for a representative 4-year realization. The parameter $w_a$ is unconstrained and is therefore not shown.}
    \label{fig:DE_w0_1D}
\end{figure}

In a more optimistic 10-year scenario, the larger number of sirens ($\sim$2.7 times; Table~\ref{tab:N_EMRI_MBHB}) tightens all constraints.
Separately, EMRIs and MBHBs reach 1.0\% and 1.5\% on $h$, and 8.6\% and 9.1\% on $\Omega_m$. 
In 1D $\Lambda$CDM runs, both populations achieve sub-percent $h$ constraints ($0.2\%$ and $0.5\%$) and percent-level $\Omega_{m}$. 
For DE, $w_0$ reaches 5.9-7.6\%, while $w_a$ remains unconstrained.

Combining populations yields $\langle \delta_h \rangle \simeq 0.6\%$, comparable to Planck+BAO under $\Lambda$CDM~\cite{Planck:2018vyg} and more constraining than typical late-time determinations (distance ladder~\cite{Riess:2021jrx} and time-delay strong lensing~\cite{TDCOSMO:2023hni}).
We obtain $\langle \delta_{\Omega_{m}} \rangle \simeq 4.7\%$ (comparable to supernova-only~\cite{Brout:2022vxf}), and $\langle \delta_{w_0} \rangle \simeq 4.9\%$, providing an independent late-time test complementary to EM probes~\cite{DESI:2025zgx} which can be used as anchor to joint analyses and improve constraints under alternative distance calibrations. 

Overall, these results show that LISA dark and bright standard sirens offer complementary redshift leverage and motivate joint analyses in which GW standard sirens provide an independent route to late-time cosmology.\\

\textit{Discussion and conclusions---}
Our EMRI results are broadly consistent with previous LISA dark siren studies~\cite{Laghi:2021pqk,liu2024probing,Liu:2025sgz}; quantitative differences mainly reflects selection thresholds and waveform/population assumptions (earlier works often adopted SNR$>$100 and/or more optimistic waveform prescriptions, improving results and precision).
Our MBHB bright siren constraints are consistent with recent forecasts under similar counterpart assumptions~\cite{Mangiagli:2023ize}.

Our cosmological posteriors are conditional on the assumed EMRI and MBHB population models used to construct the selection function, and we do not marginalize over uncertainties in the underlying population parameters.
A fully self-consistent joint inference of cosmology and populations parameters could broaden constraints, but may also gain information from redshift-dependent population features and correlations.

Our results further rely on a set of methodological assumptions that are standard in forecasting but warrant discussion. 
Concerning EMRIs, FIM uncertainties may under-represent non-Gaussianities and degeneracies that arise in full Bayesian analyses, while our dark siren treatment assumes a sufficiently complete and accurate galaxy catalog such that the true EMRI galaxy host is present. 
For MBHBs, we adopt the most optimistic EM-counterpart assumptions from~\cite{Mangiagli:2022niy}, which can underestimate the impact of collimated radio jets and dust obscuration. Furthermore, we do not model in detail host-identification challenges within the sky area (often relying on time-coincident radio flare or X-ray source; see, e.g.,~\cite{Drake:2025koq}). 
Waveform systematics, especially for EMRIs, remain an additional source of uncertainty~\cite{Babak:2017tow,Chua:2017ujo}.

At the same time, our study is conservative in several respects: we consider only a limited set of benchmark populations, do not exploit full 3D clustering information in the dark siren analysis (see End Matter), 
adopt relatively high SNR threshold (lower-SNR events could contribute as spectral sirens), and do not incorporate potential host weighting schemes that can strengthen dark siren inference. 
As LISA data-analysis methods and astrophysical modeling mature, incorporating these ingredients will be essential to propagate systematics and population uncertainties robustly.

In summary, this study demonstrates that combining dark and bright standard sirens in LISA provides complementary redshift leverage and reduces key cosmological degeneracies. This motivates continued effort toward end-to-end analyses that jointly model selection effects, populations, waveform systematics, and counterpart/host identification. In this broader context, LISA will open a new window on the GW universe and enable an independent route to late-time cosmology with systematics largely distinct from traditional EM probes.\\

\textit{Acknowledgements---}
We thank Stas Babak for useful comments and discussions and D. Izquierdo-Villalba for providing the galaxy catalog used in the EMRI dark siren analysis.

D.L. acknowledges funding from the CNES Postdoctoral Grant ``Gravitational wave cosmology with LISA'' and the UZH Postdoc Grant ``LISA Cosmo Sirens'', grant no. [K-72341-01-01]. D.L., N.T.~and S.M.~acknowledge support from the French space agency CNES in the framework of LISA. D.L. acknowledges computational resources from the CNES computing cluster \emph{TREX}.
A.S.~acknowledges support by the European Union’s H2020 ERC Advanced Grant ``PINGU'' (Grant Agreement: 101142079).
N.T.~acknowledges financial support from the Agence Nationale de la Recherche (ANR) through the MRSEI project ANR-24-MRS1-0009-01. This project has received funding from the European Union’s Horizon 2020 research and innovation program under the Marie Skłodowska-Curie grant agreement No. 101066346 (MASSIVEBAYES).
E.B. acknowledges support from the European
Union’s Horizon ERC Synergy Grant “Making Sense
of the Unexpected in the Gravitational-Wave Sky”
(Grant No. GWSky-101167314)

\textit{Data availability---}The data supporting the findings of this article are not publicly available. However, the data are available from the authors upon reasonable request.

\bibliographystyle{apsrev4-2}
\bibliography{biblio.bib}

\clearpage
\onecolumngrid

\begin{center}
\textbf{End Matter}
\end{center}

\twocolumngrid

\appendix

\begin{center}
\textbf{Single-event likelihood derivation}
\end{center}

Here we explicitly derive the single-event likelihood that enters into Eq.~\eqref{eq:posterior}, where~\cite{Laghi:2021pqk,Gair:2022zsa}
\begin{equation}
\begin{split}
    \mathcal{L}(D_i | \lambda) = & \; \frac{1}{\alpha(\lambda)} \int \! dz \int d \Omega \\
    & \times \mathcal{L}_{\rm GW}(D_i | d_L(z, \lambda), \Omega) \, p(z, \Omega| \lambda) \text{.}
\label{eq:single_ev_lk}
\end{split} 
\end{equation}
We model the single-event GW likelihood as
\begin{equation}
\label{eq:GW_lk}
\begin{split}
    \mathcal{L}_{\rm GW}(D_i | d_L(z,\lambda), \Omega) \approx \, & \mathcal{N}\left(\hat{d}_{L,i} ; \sigma_{\hat{d}_{L,i}}^2\right)\!\Big|_{d_L(z,\lambda)} 
    \\
    & \times \mathcal{N}\left(\hat{\Omega}_i;\Sigma_{\hat{\Omega}_i}\right)\!\Big|_{\Omega}\text{,}
\end{split}
\end{equation}
where $D_i = \{\hat{d}_{L,{i}}, \hat{\Omega}_i\}$ are the ``observed'' values of luminosity distance and sky location, which represent the best estimate of these two parameters from the LISA data. The notation $\mathcal{N}(\mu,\Sigma)|_x$ denotes the probability density function of a Normal distribution with mean $\mu$ and covariance $\Sigma$ evaluated at the point $x$. To obtain the simulated observed data, $D_i$, we scatter the true value $\hat{d}^{\rm true}_{L,{i}}$ from~\cite{Babak:2017tow} with an uncertainty given by
\begin{equation}
\label{eq:sigma_dl}
{\sigma_{\hat{d}_L}} = \sqrt{\left({\sigma^{\rm GW}_{\hat{d}_L}}\right)^2 + \left({\sigma^{\rm WL}_{d_L}}\right)^2}\,,
\end{equation}
where ${\sigma^{\rm GW}_{\hat{d}_L}}$ is the LISA luminosity distance error obtained in~\cite{Babak:2017tow}, while the weak lensing error ${\sigma^{\rm WL}_{d_L}}$ is modeled as in~\cite{Tamanini:2016zlh}, corrected by a factor 1/2 (see~\cite{hirata2010reducing,cusin2021characterization}); 
here we adopt a cosmology-independent weak-lensing fit (but see~\cite{Vaskonen:2026ubd} for the case where the weak-lensing magnification distribution is evaluated as cosmological parameters vary).
Since most of our MBHBs are at $z\gtrsim1$, we neglect the peculiar-velocity term in the MBHB distance error budget, which is subdominant in this regime where weak lensing dominates (see Fig. 3 of~\cite{Mangiagli:2023ize}). This contribution can be straightforwardly included in Eq.~\eqref{eq:sigma_dl}, e.g., using the fit of~\cite{Kocsis:2005vv}.
Similarly, $\hat{\Omega}_i$, the observed GW source sky position in ecliptic coordinates, is obtained by scattering $\Omega_i$ according to the 2D covariance matrix $\Sigma_{\hat{\Omega}_i}$ computed in~\cite{Babak:2017tow}.

The redshift and sky location prior $p(z, \Omega | \lambda)$ can be factorized as 
\begin{equation}
\label{eq:z_prior1}
    p(z, \Omega | \lambda) = \sum_{j=1}^{N_{\rm gal}} p_{{\rm red},j}(z|z_j, \lambda) \,  \delta^{(2)}(\Omega - \hat{\Omega}_j) \\.
\end{equation}
The term $p_{{\rm red},j}(z|z_j,\lambda)$ models the galaxy redshift prior for galaxy $j$, which assumes different forms for the EMRI (dark siren) and MBHB (bright siren) analyses, while we assume that the galaxy sky position has negligible uncertainty.

Substituting Eqs.~\eqref{eq:GW_lk} and~\eqref{eq:z_prior1} in Eq.~\eqref{eq:single_ev_lk} and integrating over angles, after defining the sky localization weight
\begin{equation}
w_{i,j}^{\rm sky} = \mathcal{N}\left(\hat{\Omega}_i ; \, \Sigma_{\hat{\Omega}_i}\right)\!\!\Big|_{\hat{\Omega}_j},
\end{equation}
we arrive at 
\begin{equation}
\begin{split}
    \mathcal{L} (D_i | \lambda) = 
     & \frac{1}{\alpha(\lambda)} \int_{z^-}^{z^+} \mathrm{d}z \; \mathcal{N}\left(\hat{d}_{L,i}; \sigma_{\hat{d}_{L,i}}^2\right)\!\Big|_{d_L(z,\lambda)} \\ 
     & \times \sum_{j=1}^{N_{\rm gal}} w_{i,j}^{\rm sky} \,  p_{{\rm red},j}(z | z_j, \lambda) \text{.}
\end{split}
\label{eqn:single_ev_lk_final}
\end{equation}
For the EMRI dark siren analysis, the galaxy redshift prior is modeled as an equal-weighted Gaussian mixture in redshift, weighted by a comoving volume factor term~\cite{Gair:2022zsa,Mastrogiovanni:2023emh,Borghi:2023opd}
\begin{equation}
\label{eq:eq:EMRI_z_prior}
\begin{split}
p_{{\rm red},j}^{\rm (EMRI)}(z | \lambda) = \frac{\mathcal{N}\!\left(z_j ;\,\sigma_j^2\right) \frac{\mathrm{d}V_c}{\mathrm{d}z}(z, \lambda)}{\int\!\mathrm{d}z' \! \sum_{j=1}^{N_{\rm gal}}\! \mathcal{N}\!\left(z_j ;\,\sigma_j^2\right)\! \frac{\mathrm{d}V_c}{\mathrm{d}z}(z', \lambda)} \,,
\end{split}  
\end{equation}
We note that this is not exactly the same as Eq.~(16) in~\cite{Gair:2022zsa}, because the latter normalizes the distribution per galaxy rather than after summing over galaxies.
In the above, the inclusion of the volume weighting gives extra weight to galaxies at higher redshifts, as these are already more abundant in the sum. The reason for this correction is that the EMRI catalogues used in~\cite{Babak:2017tow} were not uniform in comoving volume, but based on a more complex population model in which the EMRI rate evolved with redshift. We found that adding the volume weighting here gave a better match to the simulated population and unbiased results in the final inference. The fact that this correction was necessary emphasizes the need to do simultaneous population and cosmological parameter inference in the final analysis of LISA data. Such joint fitting is beyond the scope of the current paper. The results presented here can be thought of as a best case in which all population modeling uncertainties are tightly constrained.

For the MBHB bright siren analysis, the sum in Eq.~\eqref{eqn:single_ev_lk_final} collapses to a single term, while the sky localization weight is equal to one, so that
\begin{equation}
    p^{\rm (MBHB)}_{\rm red}(z|\lambda)\propto  \mathcal{N}\left(\hat{z}_{\rm EM} ;\,\sigma_{\rm EM}^2\right)\!\!\bigl|_{z} \\,
\end{equation}
where $\hat{z}_{\rm EM}$ and $\sigma_{\rm EM}$ are the measured EM counterpart redshift and its estimated instrumental uncertainty, respectively, as computed in~\cite{Mangiagli:2022niy}. 

We treat GW selection effects differently for EMRIs and MBHBs. In the EMRI analysis, due to the challenges presented in the estimation of the selection function with traditional methods (see~\cite{Chapman-Bird:2022tvu,Singh:2026sad}), we approximate the selection function $\alpha(\lambda)$ directly from the catalog of detected EMRIs as the fraction of sources that would be detectable under cosmology $\lambda$, $\alpha(\lambda) \propto N_{\rm det}(\lambda)$. Assuming that this quantity depends solely on EMRI detection through the luminosity distance (the only cosmology-related GW parameter in the likelihood), we have 
\begin{equation}
\begin{split}
    \alpha(\lambda) & \propto  \int \mathrm{d}z \, \mathrm{d}{\theta} \, p_{\rm pop}(z,\theta) \, P_{\rm det}\left(\rho(z,\theta,\lambda) > \rho_{\rm thr}\right)\\
    & \approx \frac{1}{N_{\rm sim}} \sum_i^{N_{\rm sim}} H(\rho_i(\lambda) - \rho_{\rm thr}) ,
\end{split}
\label{eqn:selection_function}
\end{equation}
where $N_{\rm sim}$ is the number of simulated EMRIs, $p_{\rm pop}(z,\theta)$ represents the population distribution at redshift $z$ described by GW parameters $\theta$, which we assume to be fairly represented by the population detected in 10 years, and $P_{\rm det}(\rho(z,\theta,\lambda) > \rho_{\rm thr})$ is the probability of detection of a GW source of parameters $\theta$ at a given $z$ assuming a cosmology $\lambda$, which we assume to only depend on the SNR $\rho$ being larger than a given threshold $\rho_{\rm thr}$.
We numerically estimate Eq.~\eqref{eqn:selection_function} by assuming an SNR scaling $\rho \propto d_L^{-1}$ and by counting how many sources pass the threshold for different cosmologies allowed by the prior $p(\lambda)$ in each of the two scenarios ($\Lambda$CDM and DE).
Inclusion of $\alpha(\lambda)$ is important to recover unbiased results, particularly when using lower SNR detection thresholds, for which the detection efficiency is a stronger function of the cosmological parameters.

For the MBHB bright siren analysis, following~\cite{Mangiagli:2023ize}, we assume $\alpha(\lambda)$ to be constant, since the detection efficiency does not appreciably change over the range of cosmological priors adopted in this analysis~\cite{Mangiagli:2023ize}.

For each event, we integrate numerically the likelihood Eq.~\eqref{eqn:single_ev_lk_final} over the redshift interval $[z^-, z^+]$ for the EMRIs (the integration boundaries are defined in the next Section), and $[\hat{z}_{\rm EM} - 5\sigma_{\rm EM}, \hat{z}_{\rm EM} + 5\sigma_{\rm EM}]$ for the MBHBs, and infer the posterior distribution Eq.~\eqref{eq:posterior} separately for dark and bright siren analyses with \texttt{cosmoLISA}~\cite{cosmolisa}. The posterior distribution is explored with \texttt{nessai}~\cite{nessai,Williams:2021qyt,Williams:2023ppp}.

\begin{center}
\textbf{Dark siren simulated data}
\end{center}

In this Section, we describe our procedure for simulating dark siren observations. This procedure is based on~\cite{Laghi:2021pqk,muttoni2023,liu2024probing} and involves cross-matching the localization error volume (LEV) of each EMRI with a galaxy light cone in redshift space. 

We use a customized light cone generated with a state-of-the-art semi-analytic evolution model~\cite{Henriques:2014sga,IzquierdoVillalba2019}, covering one octant of the sky up to $z\leq1$ and containing galaxies with $M_{*} > {10^{10}}M_{\odot}$ (see~\cite{muttoni2023}, Sec. IV A, for more details on the light cone, and~\cite{Laghi:2021pqk}, App. A1, for the impact of a lower galaxy mass threshold).

We select events from the AKS catalog of~\cite{Babak:2017tow} with SNR$>$50. 
This SNR cut ensures that the selected events have sky location uncertainties and redshifts that are smaller than the light cone's aperture and redshift upper boundary.
For each EMRI at a true luminosity distance $\hat{d}_L^{\rm true}$, we compute the uncertainty $\sigma_{\hat{d}^{\rm true}_L}$ using Eq.~\eqref{eq:sigma_dl}, assuming the true EMRI redshift and luminosity distance.
We draw an observed luminosity distance as $\hat{d}_L\sim\mathcal{N}(\hat{d}_L^{\rm true};\sigma_{\hat{d}^{\rm true}_L})$ and convert the interval $\hat{d}_L \pm 3 \sigma_{\hat{d}^{\rm true}_L}$ into a redshift interval $\left[z_{\rm fid}^-, z_{\rm fid}^+\right]$ using Eq.~\eqref{eq:dL_z} and assuming our fiducial cosmology. We then list all the galaxies in the light cone having cosmological redshift $z^{\rm cos}_{\rm gal} \in \left[z_{\rm fid}^-, z_{\rm fid}^+\right]$. 
Next we scatter the EMRI true angular position ${\hat{\Omega}}^{\rm true}$ in ecliptic coordinates by extracting a new LEV angular center of latitude $\hat{\theta}$ and longitude $\hat{\phi}$ given by $\hat{\Omega} \sim \mathcal{N}({\hat{\Omega}}_{\rm gal,tr}; \Sigma_{\hat{\Omega}^{\rm true}})$, where $\hat{\Omega}_{\rm gal,tr} = (\cos\hat{\theta}_{\rm gal, tr}, \hat{\phi}_{\rm gal, tr})$ is the true host sky location and $\Sigma_{\hat{\Omega}^{\rm true}}$ is the EMRI sky location covariance matrix from~\cite{Babak:2017tow}.
The LEV center is thus defined by the vector $(\hat{d}_L, \hat{\Omega})$. We remark that this procedure preserves the clustering properties of sky position and distance separately, while not fully exploiting the full 3D clustering information. 
Next, using Eq.~\eqref{eq:dL_z}, we find the largest possible redshift shell from the luminosity distance interval, $\hat{d}_L \pm 3 \sigma_{\hat{d}^{\rm true}_L}$, that is permitted by the cosmological prior $p(\lambda)$ defined by the $\Lambda$CDM and DE models. We denote this interval $\left[z_{\rm cp}^-, z_{\rm cp}^+\right]$. 
Finally, we further enlarge this redshift interval to account for the fact that, due to galaxy peculiar velocities, cosmological and observed galaxy redshifts will differ, $z^{\rm cos}_{\rm gal} \neq z^{\rm obs}_{\rm gal}$. Assuming $\sigma_{\rm pv} = \langle v_p \rangle/c\simeq0.0023$, where $c$ is the speed of light and $\langle v_p \rangle = 700 \text{ km s}^{-1}$ is the mean peculiar velocity estimated from the galaxy light cone, we define the galaxy redshift uncertainty as
\begin{equation}
\label{eq:sigma_z}
    \sigma_z (z) = \sigma_{\rm pv}(1+z)\,.
\end{equation}
Using Eq.~\eqref{eq:sigma_z}, we compute the final LEV redshift boundaries as $z^{\pm} = z_{\rm cp}^{\pm} \pm \sigma_{z}(z_{\rm cp}^{\pm})$, checking that $z^+ \leq 1$, otherwise a new true host is drawn and the whole procedure is repeated. 
Finally, we add galaxy hosts to the LEV by selecting all the galaxies with \emph{observed} redshift $z^{\rm obs}_{\rm gal} \subset \left[z^-, z^+\right]$ and 
sky position within the 3$\sigma$ sky-localization ellipse;
to reduce computational cost when building the redshift-prior interpolant, we restrict to this subset.
In summary, for each galaxy in the LEV, we use the observed redshift $z^{\rm obs}_{\rm gal}$, the redshift uncertainty $\sigma_z(z^{\rm obs}_{\rm gal})$,  Eq.~\eqref{eq:sigma_z}, to compute the galaxy redshift prior, Eq.~\eqref{eq:eq:EMRI_z_prior}.

\end{document}